\documentstyle[epsf]{mn}

\newif\ifAMStwofonts
\ifoldfss
  \ifCUPmtlplainloaded \else
    \NewTextAlphabet{textbfit} {cmbxti10} {}
    \NewTextAlphabet{textbfss} {cmssbx10} {}
    \NewMathAlphabet{mathbfit} {cmbxti10} {} % for math mode
    \NewMathAlphabet{mathbfss} {cmssbx10} {} %  "   "    "
  \fi
  \ifAMStwofonts
    \ifCUPmtlplainloaded \else
      \NewSymbolFont{upmath} {eurm10}
      \NewSymbolFont{AMSa} {msam10}
      \NewMathSymbol{\upi}     {0}{upmath}{19}
      \NewMathSymbol{\umu}     {0}{upmath}{16}
      \NewMathSymbol{\upartial}{0}{upmath}{40}
      \NewMathSymbol{\leqslant}{3}{AMSa}{36}
      \NewMathSymbol{\geqslant}{3}{AMSa}{3E}

      \let\leq  \leqslant 
       
    \fi
  \fi
\fi % End of OFSS

\ifnfssone
  \newmathalphabet{\mathit}
  \addtoversion{normal}{\mathit}{cmr}{m}{it}
  \addtoversion{bold}{\mathit}{cmr}{bx}{it}
  \newmathalphabet{\mathbfit} % math mode version of \textbfit{..}
  \addtoversion{normal}{\mathbfit}{cmr}{bx}{it}
  \addtoversion{bold}{\mathbfit}{cmr}{bx}{it}
  \newmathalphabet{\mathbfss} % math mode version of \textbfss{..}
  \addtoversion{normal}{\mathbfss}{cmss}{bx}{n}
  \addtoversion{bold}{\mathbfss}{cmss}{bx}{n}
  \ifAMStwofonts
    \ifCUPmtlplainloaded \else
and
your
      \UseAMStwoboldmath
      \makeatletter
      \new@mathgroup\upmath@group
      \define@mathgroup\mv@normal\upmath@group{eur}{m}{n}
      \define@mathgroup\mv@bold\upmath@group{eur}{b}{n}
      \edef\UPM{\hexnumber\upmath@group}
      \new@mathgroup\amsa@group
      \define@mathgroup\mv@normal\amsa@group{msa}{m}{n}
      \define@mathgroup\mv@bold\amsa@group{msa}{m}{n}
      \edef\AMSa{\hexnumber\amsa@group}
      \makeatother
      \mathchardef\upi  "0\UPM19
      \mathchardef\umu  "0\UPM16
      \mathchardef\upartial  "0\UPM40
      \mathchardef\leqslant  "3\AMSa36
      \mathchardef\geqslant  "3\AMSa3E

      \let\leq  \leqslant 

    \fi
  \fi
\fi % End of NFSS release 1

\ifnfsstwo

  \def\spose#1{\hbox to 0pt{#1\hss}}
  \def\lta{\mathrel{\spose{\lower 3pt\hbox{$\mathchar"218$}}
     \raise 2.0pt\hbox{$\mathchar"13C$}}}
  \def\gta{\mathrel{\spose{\lower 3pt\hbox{$\mathchar"218$}}
     \raise 2.0pt\hbox{$\mathchar"13E$}}}
  \DeclareMathAlphabet{\mathbfit}{OT1}{cmr}{bx}{it}
  \SetMathAlphabet\mathbfit{bold}{OT1}{cmr}{bx}{it}
  \DeclareMathAlphabet{\mathbfss}{OT1}{cmss}{bx}{n}
  \SetMathAlphabet\mathbfss{bold}{OT1}{cmss}{bx}{n}
  \ifAMStwofonts
    \ifCUPmtlplainloaded \else
      \DeclareSymbolFont{UPM}{U}{eur}{m}{n}
      \SetSymbolFont{UPM}{bold}{U}{eur}{b}{n}
      \DeclareSymbolFont{AMSa}{U}{msa}{m}{n}
      \DeclareMathSymbol{\upi}{0}{UPM}{"19}
      \DeclareMathSymbol{\umu}{0}{UPM}{"16}
      \DeclareMathSymbol{\upartial}{0}{UPM}{"40}
      \DeclareMathSymbol{\leqslant}{3}{AMSa}{"36}
      \DeclareMathSymbol{\geqslant}{3}{AMSa}{"3E}

      \let\leq  \leqslant 

    \fi
  \fi
\fi % End of NFSS release 2

\ifCUPmtlplainloaded \else
  \ifAMStwofonts \else % If no AMS fonts
    \def\upi{\pi}
    \def\umu{\mu}
    \def\upartial{\partial}
\fi

\title{Stellar models and Hyades: the Hipparcos test}

\author[V. Castellani et al.]
       {V.Castellani$^{1,2}$, S. Degl'Innocenti$^{1,2}$, P. G. Prada  
Moroni$^{3,4}$\\
  $^1$ Dipartimento di Fisica, Universit\`a di Pisa, piazza Torricelli  
2, 56126 Pisa, Italy \\
  $^2$ INFN Sezione di Pisa, via Livornese 1291, 56010 S. Piero a Grado,  
Pisa, Italy \\
  $^3$ Dipartimento di Fisica, Universit\`a di Genova, via Dodecaneso  
33, 16146 Genova, Italy \\
  $^4$ INFN Sezione di Genova, via Dodecaneso 33, 16146 Genova, Italy \\
}

\pagerange{\pageref{firstpage}--\pageref{lastpage}}
\pubyear{ 00}

\begin{document}

\maketitle

\label{firstpage}

\baselineskip 20pt

\begin{abstract}

We compare theoretical stellar models for Main Sequence (MS) stars
with the Hipparcos database for the Hyades cluster to give a warning
against the uncritical use of available theoretical scenarios and to
show how formal MS fittings can be fortuitous if not fictitious.
Moreover, we find that none of the current theoretical scenarios
appears able to account for an observed mismatch between theoretical
predictions and observations of the coolest Hyades MS stars. Finally,
we show that current theoretical models probably give too faint He
burning luminosities unlike the case of less massive He burning
models, with degenerate progenitors, which have been suggested to
suffer the opposite discrepancy.

\end{abstract}


\section{Introduction}  

The prediction of cluster isochrones is one of the most popular
results of stellar evolutionary theories and in the current literature 
the comparison of
theoretical isochrones with observed CM diagrams of stellar
clusters is widely adopted to
investigate the evolutionary status of cluster stars,  
yielding information on cluster distances and  
ages. However, the success achieved by theory in reproducing all the main
evolutionary features observed in actual clusters has perhaps produced 
too much confidence in the quantitative theoretical predictions,
which are often taken at their face values without accounting for the
uncertainties still existing in the theoretical scenario. One
should indeed bear in mind that theoretical predictions are still affected by
uncertainties due either to the physics input (equation of state, EOS,
opacity, etc.)  or to assumptions about the efficiency of  
some macroscopic mechanisms like core overshooting, diffusion,  
superadiabatic convection and so on.   

Moreover, the fitting of theoretical isochrones to cluster stars does
depend on the adopted transformations from the theoretical to the
observational plane, often with the additional degrees of freedom
introduced by current uncertainties in the cluster distance and reddening.  
The different varied results in the recent literature concerning
theoretical predictions and, in turn, the evaluation of evolutionary
parameters for a given cluster can be taken as a clear indication of
the uncertainty in this kind of procedure.  On this basis
evolutionary theory is just indicative whereas experiments
(i.e. observations) must provide  the right answer.

%%%%%%  FIGURA  1

\begin{figure}
\centerline{\epsfxsize=  8 cm \epsfbox{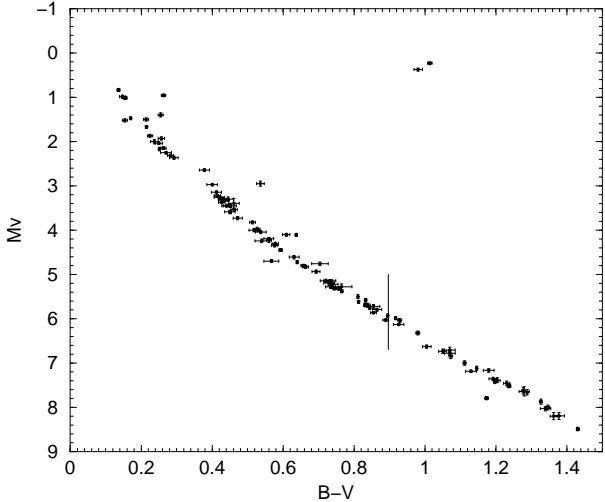}}
\caption{The CMD for the Hyades, using the subset of Dravins et
al. (1997), and Madsen, Dravins, \& Lindegren (2000) data  
where known spectroscopic binaries were also excluded.
The vertical line marks the right boundary of the region fitted
by Perryman et al. 1998.)
}
\end{figure}

In this context the results of the Hipparcos satellite have opened a new
era in stellar astrophysics, fixing the distance and -thus- the
absolute magnitude of stars in several nearby open clusters,
removing a noisy degree of freedom affecting previous
evaluations. Among these clusters we will focus our attention on the
Hyades cluster, which  has already been the object of several
careful investigations and, in particular, for which one has reliable
evaluations of the cluster metallicity together with indications for a
rather negligible reddening (see e.g. Perryman et al. 1998).  
In this paper we will take advantage of the beautiful CM diagram for  
cluster  members presented by Dravins  
et al. (1997) and Madsen, Dravins, \& Lindegren (2000),   
reported here in Fig.1.
MS stars hotter than B-V$\sim$0.9 have already been 
compared with suitable theoretical models by Perryman et al.  
(1998). However, the well 
defined sequence of MS stars gives the opportunity  
for testing much cooler stellar models than the above quoted limit.

Moreover, one finds that the diagram provides evidence for
two red giants, easily interpreted as He burning structures,  
worthy of comparison with the predictions of current  
evolutionary scenarios. This paper deals with    
these questions. For this purpose, the next section
will first discuss the level of confidence for
theoretical predictions. On this basis, we will use Hyades  
stars as a test of several current evolutionary scenarios.

\section{The Hyades MS}

Present stellar models were computed by adopting our version of the FRANEC
evolutionary code (Chieffi \& Straniero 1989; Ciacio, Degl'Innocenti,
Ricci 1997), which relies on the most recent input physics (Cassisi et
al. 1998). In particular, to start our investigation we adopted the
OPAL equation of state (Rogers et al. 1996) using model atmospheres by
Kurucz (1993) to transform theoretical results into the observational
plane (V, B-V).
 
On these grounds, we computed suitable canonical (without core
overshooting) isochrones, assuming for the Hyades stars Z=0.024 (see
e.g. Perryman et al. 1998) together with Y=0.278, as given by
extrapolation of the linear relation between Y and Z, connecting metal
poor Pop.II stars ($Z={10}^{-4}$ $Y=0.23$) to the results of standard
solar models ($Z=0.02$ $ Y=0.27$) (see e.g. Pagel \& Portinari 1998,
Castellani, Degl'Innocenti, Marconi 1999).

%FIGURA 2

\begin{figure}   
\label{alfa1-2.2}
\centerline{\epsfxsize=  8 cm \epsfbox{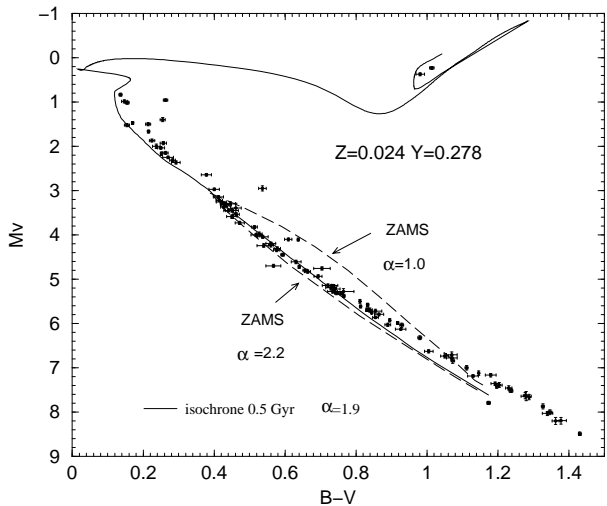}}
\caption{CM diagram of Hyades as in Fig.1 compared with a 0.5 Gyr theoretical
isochrone for the Hyades composition (Z=0.024, Y=0.278) and
$\alpha$=1.9, and with Zero Age Main Sequences (ZAMSs) for the same chemical composition and two
extreme values of the mixing length: $\alpha$ =1 and $\alpha$ =2.2.
Equation of state is from Livermore (Rogers et al. 1996).
Colour transformations and bolometric corrections are from Kurucz (1993).}
\end{figure}

As is well known, an exaustive theory of convection in the turbulent
external regions of stars is not yet available; thus the Mixing Length
Theory (MLT) is generally adopted, where the free parameter $\alpha$
is varied to tune the efficiency of superadiabatic convection until
the agreement with the observations is reached.  As a consequence, stellar
evolution cannot give firm predictions about the effective temperature
and the radius of cool stars with convective envelopes, which depend
on the assumptions regarding the value of $\alpha$. To have a look into
such an occurrence, Fig.2 shows again the CM diagram of the Hyades stars
but comparing them with MS loci computed under the different labelled
assumptions about the value of the free parameter $\alpha$.  One
easily recognizes that, within a given theoretical scenario, theory
can give firm predictions only for stars hotter than B-V$\sim$0.4
(where convection vanishes) or cooler than B-V$\sim$1.2 (where
convection becomes adiabatic).

%FIGURA 3

\begin{figure}
\centerline{\epsfxsize= 8 cm \epsfbox{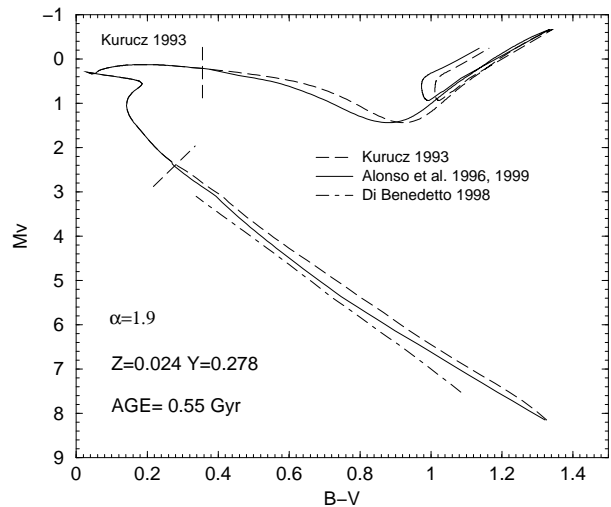}}
\centerline{\epsfxsize= 8 cm \epsfbox{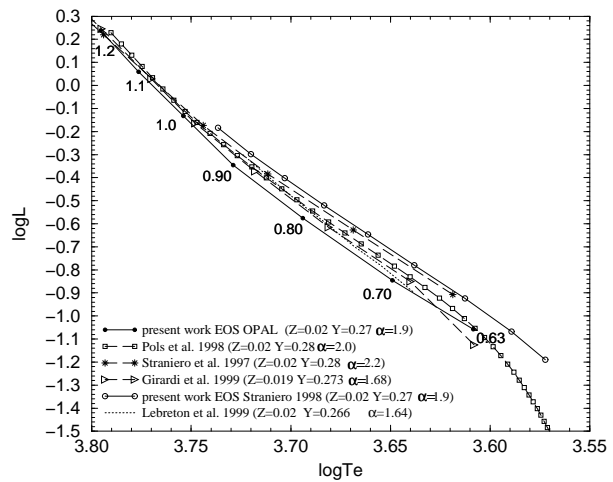}}
\caption{Upper panel: 0.55 Gyr theoretical isochrone (Z=0.024, 
Y=0.278, $\alpha=1.9$) transposed in the observational plane by using
the labelled colour trasformations.
Lower panel:ZAMS location in the HR diagram from the labelled
papers. ZAMS from the present work with two different adopted Equations of State (EOS) are
drawn with a solid line (see text).
Mass values at given positions on a selected ZAMS are also shown. 
The luminosity is expressed in solar units.
}
\label{colori}
\end{figure}

In this context, one finds that at the hotter end of the ``uncertainty
region" (0.4$<$B-V$<$1.2) theoretical predictions appear to be in
reasonable agreement with observation. Luckily enough, below
B-V$\approx$0.4 the observational sequence is well within the
uncertainty region, and the figure discloses that the assumption
$\alpha$=1.9 fits the observations best, at least down to
B-V$\sim$0.8.  For B-V$>$0.8 models with $\alpha$=1.9 become bluer
than observations but up to B-V$\approx$1.1 this problem can be
overcome by decreasing the mixing length parameter.  However, at the
cooler end, theoretical predictions give models that are too blue,
independently of the assumption on the mixing length, revealing that
something is wrong, either in the models or in the adopted colour
transformations.

However, it is not difficult to recognize that even the above quoted
agreement is far from being firmly established. Fig.3 (upper panel)
shows a 0.55 Gyr theoretical isochrone transposed in the observational
plane by using different colour trasformations available in
present-day literature, as provided by Alonso et al. (1996, 1999) or
by Di Benedetto (1998).  As marked in the same figure, in the hotter
portion of the diagram relations of Alonso et al. have been
implemented with Kurucz (1993) model atmospheres shifted so as to
match the Alonso et al. ones in their hotter limit of validity. Di
Benedetto's relations are available presently only for MS stars.

One finds that improved semiempirical colours by Alonso et al. (1996,
1999) would make our theoretical MS bluer, whereas the recent
empirical colours by Di Benedetto (1998) will make it even bluer. In
other words the stellar models become, for each given MS colour,
underluminous. By adopting, as many people do, Alonso et al. colours
the agreement between theory and observation, inside the ``uncertainty
region'' of Fig.2, could still be reached by tuning the mixing length
parameter while a variation of $\alpha$ cannot solve the increasing
disagreement below B-V$\approx$ 1.1.  With the choice of Di Benedetto
colours there are no reasonable assumptions for the efficiency of the
external convection which could reconcile theory and observation for
stars both hotter than B-V$\approx$0.5 or cooler than B-V$\approx$1.0.
We conclude that, for any choice of mixing length and colour
transformations, stellar models are certainly underluminous for
colours redder than B-V$\approx$1.1.

Luckily enough brighter models are within the range of current theories. As shown  
in the same figure (lower panel) one finds in the  
literature that MS in all cases computed with ``reasonable" input
physics cover a non negligible range of luminosity, all being
brighter than the MS computed with ``the most updated" physics.
Among the brightest, one finds the MS by Straniero, Chieffi, Limongi (1997),
which differs mainly from our computations  
by adopting the Equation of State by Straniero
1998 (see also Straniero 1988). As a matter of fact, and as shown in the same figure,  
our models computed, however,  with the above quoted equations of state  
 closely follow the results of Straniero et al. (1997).

%FIGURA 4

\begin{figure}
\centerline{\epsfxsize= 8 cm \epsfbox{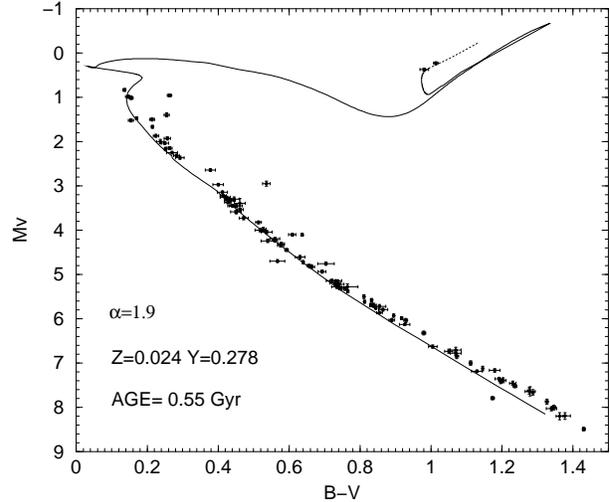}}
\caption{0.55 Gyr theoretical isochrone for Z=0.024 ,Y=0.278,
$\alpha=1.9$ with Straniero 1998 EOS and Alonso et al. (1996,1999)
colour transformations (completed with Kurucz 1993 colour-temperature relations
as shown in Fig.3) compared to the observational data of Fig.1.
The dotted region of the theoretical isochrone indicates the fastest
part of the central He burning phase which is expected to be populated very little.}
\label{fit2EOS}
\end{figure}

Figure 4 reveals that the simultaneous inclusion of Straniero's EOS
and Alonso colours again provides a reasonable fitting of the
Hipparcos data, also improving the fitting for cooler stars, with only
a residual discrepancy at the cooler end of the sequence. However,
comparison of Fig.4 with the upper panel of Fig.3, shows that if the
Di Benedetto empirical colours will be confirmed, one cannot escape
the conclusion that theories should give brighter models than those
provided by current evaluations. An increase in the adopted original
He content could help in reaching a good fit all along the major
portion of the MS, but with the lower end still requiring some
modification of the adopted theoretical scenario.  However we feel
that any attempt to theoretically constrain the Hyades He content
cannot give reliable results until firm evaluations of the
colour-temperature relations will become available.

\section{Advanced evolutionary phases}

Even a quick inspection of the data in Fig.1 reveals that when including
only highly probable members, with the exclusion of binaries, the luminous
termination of the cluster MS appears rather poorly defined, making it
difficult a clear identify the Turn-Off (TO) and, thus, a precise
evaluation of the cluster age. We will discuss this point by relying on
the theoretical scenario adopted in Fig.4, as based on Straniero's EOS
and Alonso et al. colours.

%FIGURA 5

\begin{figure}
\centerline{\epsfxsize= 8 cm \epsfbox{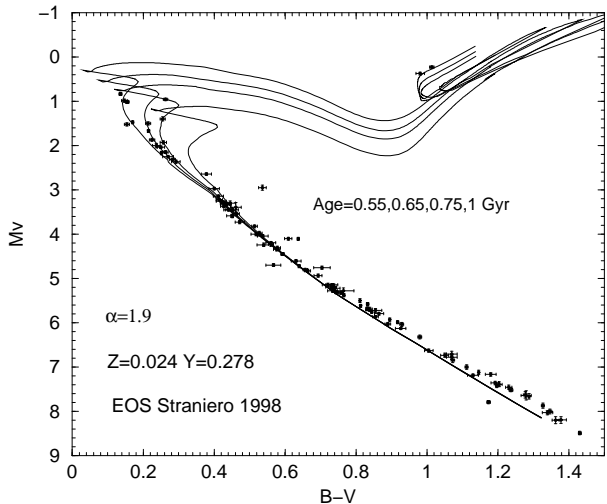}}
\caption{Theoretical isochrones for the labelled ages (Z=0.024,
Y=0.278, $\alpha=1.9$) with Straniero 1998 EOS and Alonso et
al. (1996,1999) colour transformations compared to the observational
data of Fig.1. The younger are the isochrones the brighter is the
subgiant phase and the bluer is the central He burning region.}
\label{fit2EOS}
\end{figure}

The evaluation of the cluster age depends on the choice made about the
actual Turn-Off. Oddly enough, Fig.5 shows - as an extreme case - how
four different assumptions about the cluster age are needed to cover
all the stars observed at the top of the cluster MS. If, as the most
reasonable choice, one fits the five hottest stars, then one derives for
the cluster an age of the order of 0.55 Gyr. By the way, any attempt
to detect possible signatures of core overshooting mechanisms (see
e.g. Maeder 1975, Bressan et al. 1981) appears beyond any actual
possibility.

As already mentioned, observations support the evidence for two stars in
the He burning phase, at around B-V$\approx$ 1.0.  According
to the fit with a 0.55 Gyr isochrone one expects at the cluster TO
stars as massive as 2.5 M$_{\odot}$ whereas the He burning clump would
be populated by 2.55 M$_{\odot}$ stars.  However, comparison with
predictions concerning the major phase of central He burning, as given
in Fig.5, suggests that theory is slightly
underestimating the actual star luminosities. Even if the statistics is
rather poor, this appears to be a relevant outcome since for less massive
stars with degenerate progenitors, theoretical models have been
already suggested to be, on the contrary, overluminous (see e.g. Pols
et al. 1998 and Castellani et al. 2000).

% FIGURA 6

%\vspace{1cm}
\begin{figure}
\centerline{\epsfxsize= 8 cm \epsfbox{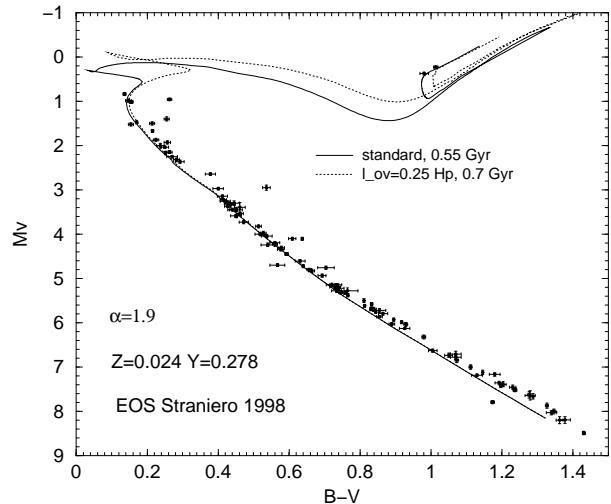}}
\caption{ Theoretical isochrones computed with: 1) no overshooting for
an age of 0.55 Gyr, 2) mild overshooting for an age of 0.7 Gyr. The
linear extension (${l}_{ov}$) of the overshooting region (out of the
zone in which the Schwarzschild criterion is fulfilled) is expressed
in terms of the pressure scale height ${l}_{ov}=\beta {H}_{p}$. For
our mild overshooting calculations we adopted $\beta = 0.1$ for
1.0M$_{\odot}$$\leq$M$\leq$1.2M$_{\odot}$ and $\beta = 0.25$ for
higher masses.  In both cases we adopted colours of Alonso et
al. (1996,1999) and Straniero (1998) EOS.  }
\label{fitovershoot}
\end{figure}

\noindent
Comparison of Fig.2 with Fig. 4 shows that the problem cannot be
overcome by adopting the OPAL EOS.  Nor the assumption of an
efficient overshooting is of actual help in solving the observed
discrepancy. This is shown in Fig.6, where we present the best fit for
cluster stars as alternatively obtained from canonical or mild
overshooting models.

% FIGURA 7

%\vspace{1cm}
\begin{figure}
\centerline{\epsfxsize= 8 cm \epsfbox{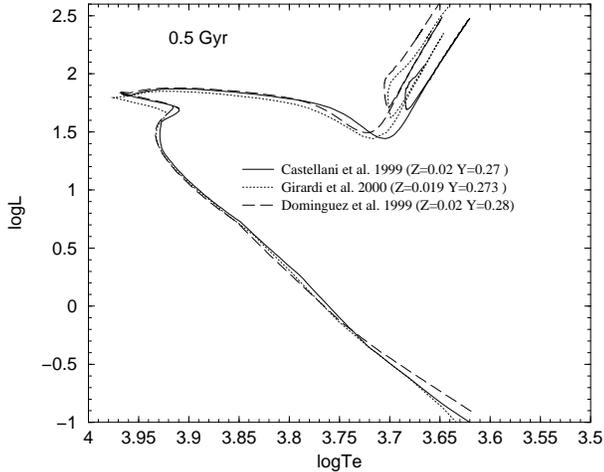}}
\caption{ 
Theoretical isochrones for 0.5 Gyr computed by:
Castellani et al. 1999 (solid line), Girardi et al. 2000 (dotted line)
and Dominguez et al. 1999 (dashed line). The luminosity is in solar units.}
\label{}
\end{figure}

\noindent
More in general, it appears that the discussed underluminosity is a
problem concerning several recent models. We already found that Pols
et al. (1998) give He burning luminosities in good agreement with
present computations (see Fig.1 in Degl'Innocenti et al. 2000).
Figure 7 discloses that similar luminosities can be also found in other
independent investigations.

Thus, at least for this cluster, one finds that
theoretical predictions for masses larger than the critical mass for
the red giant (RG) transition, i.e. stars which during their RG
evolution were not affected neither by electron degeneracy nor by
relevant neutrino cooling, seems to show the problem opposite to that
 found for He burning stars with degenerate progenitors.

\section{Conclusion}

In this paper we discussed Hipparcos data for the Hyades
cluster, showing that the fitting of theoretical isochrones to open
clusters is still affected by non negligible uncertainties in
the theoretical predictions which require a much firmer
evaluation of colour temperature relations. However, none
of the current theoretical scenarios appears able to account  
for the mismatch between theoretical predictions and observations of
the Hyades coolest MS stars. Finally we found that Hyades He burning  
stars suggest an underluminosity of the current theoretical models,  
unlike the case of less massive He burning models, with degenerate  
progenitors, which probably suffer the opposite problem.

\section{Acnowledgements}

It is a pleasure to thank A. Alonso for providing us with the Alonso
et al. colour transformations before publication and L. Lindegren for
making the corrected Hipparcos data for the Hyades available to us. We
are grateful to J.C. Mermilliod and to the referee, C. Tout, for
useful comments.  We warmly thank L. Lovitch and G. Bono for a careful
reading of the manuscript.  This work is partially supported by the
Ministero dell'Universit\`a della Ricerca Scientifica (MURST) within
the ``Stellar Evolution'' project (COFIN98).

\label{lastpage}


\begin{thebibliography}{99}

\bibitem{Alon} Alonso A., Arribas S., Martinez-Roger C., 1996, A\&A 313,  
873
\bibitem{} Alonso A., Arribas S., Martinez-Rogers C., A\&A Suppl., 1999, 140, 261
\bibitem{} Bressan A., Bertelli G., Chiosi C., 1981, A\&A 102, 25
\bibitem{} Cassisi S., Castellani V., Degl'Innocenti S. Weiss A. 1998  
A\&A Suppl., 129, 267
\bibitem{} Castellani V., Degl'Innocenti S.,  
Marconi M., 1999, A\&A 349, 834
\bibitem{} Castellani V., Degl'Innocenti S., 1999, A\&A 344, 97
\bibitem {} Castellani V., Degl'Innocenti S., Girardi L., Marconi M., Prada  
Moroni P.G., Weiss A.,2000, A\&A, 354, 150
\bibitem{} Chieffi A., Straniero O., 1989, ApJS 71, 47
\bibitem{} Ciacio F., Degl'Innocenti S., Ricci B., 1997, A\&AS 123, 449   
\bibitem{} Degl'Innocenti S., Castellani V., Girardi L., Marconi M., Prada Moroni P.G.,
           Weiss A., 2000, in ``Massive Stellar Clusters'', ASP Conference Series, vol. 211,
	   pag. 169, A. Lancon and C.M. Boily eds.
\bibitem{} Di Benedetto G. P., 1998, A\&A 339, 858
\bibitem{} Dominguez I., Chieffi A.,  Limongi M., Straniero O., 1999, ApJ 524, 225
\bibitem{} Dravins D., Lindegren L., Madsen S., Holmberg J., 1997, ESA,  
SP-402, p. 733
\bibitem{} Girardi L., Bressan A., Bertelli G., Chiosi C., 2000,
	 A\&AS 141, 371
\bibitem{} Kurucz R. L., 1993, International Astronomical Union,  
Colloquium No.138, Trieste, Italy
\bibitem{} Lebreton Y., Perrin M.N., Cayrel R., Baglin A., Fernandes J.,  
1999, A\&A 350,587  
\bibitem{} Maeder A., 1975, A\&A 40, 303
\bibitem{} Madsen S., Dravins D., Lindegren L., 2000, ApJ, in  
preparation
\bibitem{}  Pagel B.E.J., Portinari L., 1998, MNRAS 298, 747  
\bibitem{Perr} Perryman M.A.C. et al., 1998, A\&A 331, 81   
\bibitem{}  Pols O.R.,Schroeder K-P, Hurley J.R., Tout C.A., Eggleton  
P.P., 1998, MNRAS 298, 525
\bibitem{} Rogers F. J., Swenson F. J., Iglesias C. A., 1996, ApJ 456,  
902
\bibitem{} Straniero O., Chieffi A., Limongi M., 1997, ApJ 490, 425 
\bibitem{} Straniero O., 1998 (private comunication)
\bibitem{} Straniero O., 1988, A\&AS 76, 157
\end{thebibliography}
\end{document}